\documentclass[reprint, superscriptaddress]{revtex4-1}
\usepackage{array}
\usepackage{booktabs} 
\usepackage{amsmath}   
\usepackage{listings}
\usepackage{graphicx}
\usepackage{dcolumn}
\usepackage{bm}
\usepackage{braket,algorithm}
\usepackage{algorithmic}

\usepackage[colorlinks,linkcolor=blue,anchorcolor=blue,citecolor=blue,urlcolor=blue]{hyperref}
\usepackage{tikz}
\usepackage{booktabs,multirow,mathrsfs,amssymb,amsmath,amsthm}
\usepackage{caption}
\usepackage{makecell}
\usepackage{qcircuit}
\usepackage{float}

\usepackage{newunicodechar}

\newunicodechar{，}{,}

\begin{document}

\title{Quantum Semantic Communication Beyond the Shannon-Wyner Channel Capacity}

\author{Min Wang}
\thanks{These authors contributed equally to this work.}
\affiliation{Beijing Academy of Quantum Information Sciences, Beijing 100193, China}

\author{Gui-Fa Zhu}
\thanks{These authors contributed equally to this work.}
\affiliation{College of Physics and Technology, Guangxi Normal University, Guilin 541004, China}

\author{Guo-Fei Long}
\thanks{These authors contributed equally to this work.}
\affiliation{Beijing TGW Quantum Technology Co.,LTD, Beijing 100015，China}

\author{Jianxing Guo}
\affiliation{Beijing Academy of Quantum Information Sciences, Beijing 100193, China}

\author{Yu-Chen Liu}
\affiliation{Department of Physics, State Key Laboratory of Low-Dimensional Quantum Physics, Tsinghua University, Beijing 100084, China}

\author{Dong Pan}
\affiliation{Beijing Academy of Quantum Information Sciences, Beijing 100193, China}

\author{Li-Ping Nong}\email{nongliping@gxnu.edu.cn}
\affiliation{College of Physics and Technology, Guangxi Normal University, Guilin 541004, China}
\affiliation{School of Information and Communication, Guilin University of Electronic Technology, Guilin 541004, China}

\author{Gui-Lu Long}\email{gllong@tsinghua.edu.cn}
\affiliation{Beijing Academy of Quantum Information Sciences, Beijing 100193, China}
\affiliation{Department of Physics, State Key Laboratory of Low-Dimensional Quantum Physics, Tsinghua University, Beijing 100084, China}
\affiliation{Frontier Science Center for Quantum Information, Beijing 100084, China}


\maketitle



Quantum Secure Direct Communication (QSDC)—a paradigm-shifting breakthrough in quantum communication—exploits quantum states for unmediated information transmission \cite{Long02,Pan24}. Rooted in the inviolable fundamental laws of quantum mechanics, QSDC enables ultra-sensitive detection of even the faintest eavesdropping attempts, guaranteeing true communication security solely when no interference exists. If eavesdropping or intrusion is detected mid-transmission, the system instantly alerts users and severs data flow, shielding them from unauthorized tracking and mitigating hacker threats. Over two decades, QSDC has seen extraordinary advancements, currently attaining kilobit-per-second transmission over 100 km of commercial optical fiber \cite{Hu16,Zhang17,Zhu17,Qi19,Qi21,Zhang22,Wang23,Pan25,Yang25,WangNSR25}. However, its practical scalability remains constrained by insufficient transmission rates—a critical bottleneck \cite{Wyner75,Qi19}. Semantic communication, which drastically boosts transmission efficiency by extracting core information features, nevertheless stays vulnerable to malicious intrusions \cite{Qin21,Xie21,Yang22}. Integrating these paradigms promises to simultaneously enhance equivalent data rate and security \cite{PanLong25}. Herein, we propose and experimentally validate a quantum semantic communication scheme, applying it to 3D point clouds. It achieves a 46.30-fold efficiency gain over direct transmission, surpassing both Wyner and Shannon capacity limits. This breakthrough not only clears the path for large-scale QSDC deployment but also marks a pivotal milestone in quantum information science.

\section*{The Quantum Semantic Communication Framework}

\begin{figure}[htbp]
	\centering
	\includegraphics[width=0.5\textwidth]{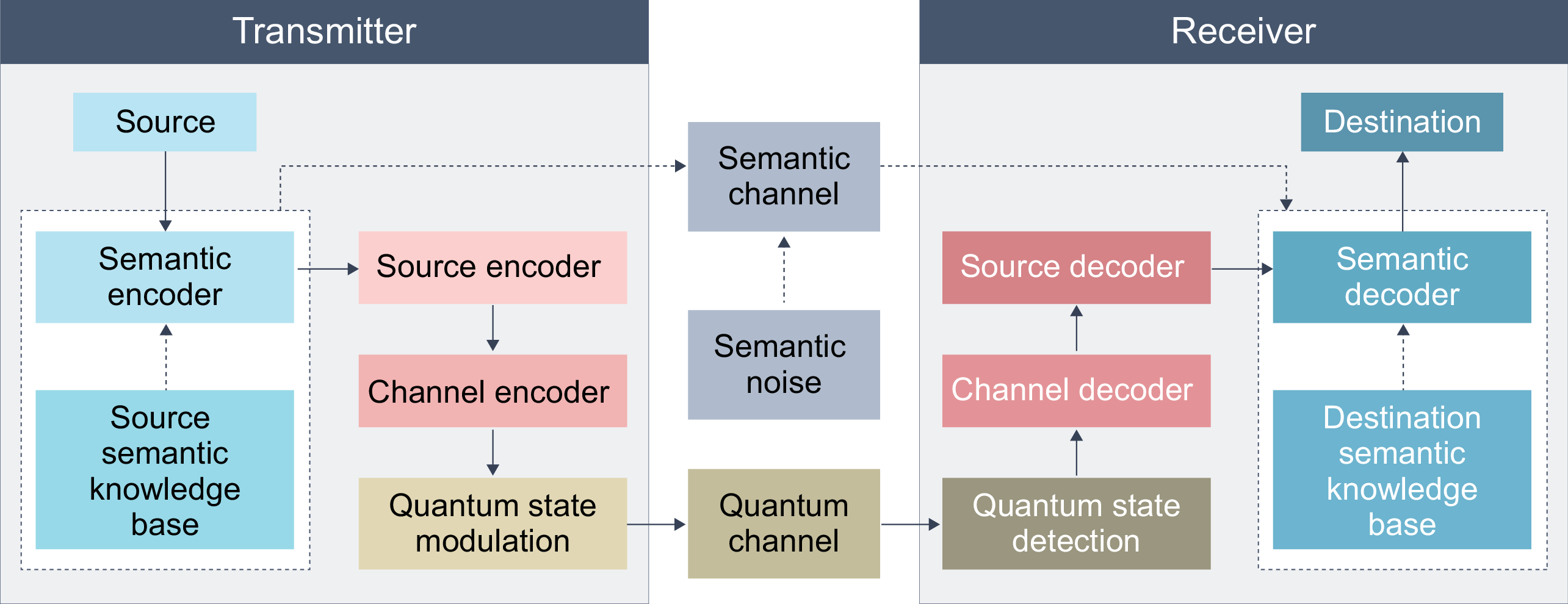}
	\caption{General framework for quantum semantic communication. }
    \label{fig1}
\end{figure}

In quantum semantic communication, the communication chain comprises a transmitter, a quantum channel, and a receiver, as illustrated in Fig.~\ref{fig1}. The communication process begins with the information source, which generates raw data (including point clouds, speech, and text). At the transmitter, the semantic encoder processes the input data by leveraging the source semantic knowledge base: this knowledge base provides critical semantic priors and contextual insights, enabling the encoder to distill a compact yet highly meaningful semantic representation.

Once the semantic representation is derived, the source encoder further converts it into codewords or symbol sequences optimized for digital processing and transmission, reducing redundant data while ensuring format uniformity. The channel encoder then embeds redundancy into these symbol sequences to enhance error resilience; subsequently, the sequences are mapped to quantum states and transmitted via QSDC technology. Crucially, prior to the transmission of these encoded quantum states, the quantum channel undergoes pre-authentication \cite{Wang24}, laying a secure foundation for the entire communication process.

During transmission, the quantum channel is subject to physical-layer noise (e.g., decoherence, phase jitter, and photon loss), which distorts the transmitted quantum states. Meanwhile, the communication process is influenced by the semantic channel, which models the distortions at the semantic level. In the context of the source and destination semantic knowledge bases, such distortions—referred to as semantic noise—may include random erasure of semantic content, incorrect substitution of semantic features, or overall semantic blurring and drift. Semantic noise typically arises from contextual mismatches, inconsistencies between the source and destination knowledge bases, or inference errors, making it challenging for the receiver to accurately recover the intended semantics.

At the receiver, the incoming quantum states are first detected and measured to obtain symbol sequences. The channel decoder applies error correction and recovery algorithms to reconstruct symbols that closely approximate the transmitted ones. The source decoder then converts the recovered symbols back into the semantic representation. With the support of the destination semantic knowledge base, the semantic decoder interprets and reconstructs the semantics: the knowledge base provides contextual priors and structural information, allowing the decoder to compensate for missing elements caused by semantic noise, resolve ambiguities, and restore an output that closely matches the original data. Finally, the system generates decoded content usable by humans or higher-level tasks, and a feedback mechanism can be employed to return confidence levels or error information to the transmitter for updating the source and destination knowledge bases and optimizing transmission performance.

3D point clouds~\cite{Wehr99,Qi17,Hackel16}, as an essential data format capable of accurately capturing spatial geometric structures, have gradually emerged as a core modality for future information interaction and intelligent perception. However, the high dimensionality and sparse distribution of point clouds render quantum communication for such data impracticable at present, which is fundamentally constrained by the Shannon-Wyner capacity, given current technological capacities, despite the fact that QSDC itself provides high-grade security guarantees. By contrast, classical semantic communication has demonstrated significant advancements in communication efficiency and robustness for 3D point clouds, while it also exhibits promising application prospects in multimodal data transmission, including scenarios such as images~\cite{Sun24}, speech~\cite{Tian24}, and text~\cite{Wang25}. Deep learning models~\cite{Chen22} have been proposed to extract essential features from point clouds, achieving effective reconstruction and recognition with minimal data transmission and greatly alleviating bandwidth constraints. However, the classical communication channels used in this approach still suffer from security vulnerabilities, being susceptible to eavesdropping and attacks~\cite{Gisin02}. 

Combining point cloud semantic communication with QSDC, we construct a quantum semantic communication of point clouds (QSCPC) scheme, which leverages the inherent security of quantum communication to ensure the secure transmission of semantic data, while employing semantic encoding and decoding to achieve efficient secure communication. This integration is particularly advantageous in scenarios with stringent security and efficiency requirements, such as autonomous driving, smart cities, and military communications.

\section*{Experiment}

\begin{figure}[t]
	\centering
	\includegraphics[width=0.5\textwidth]{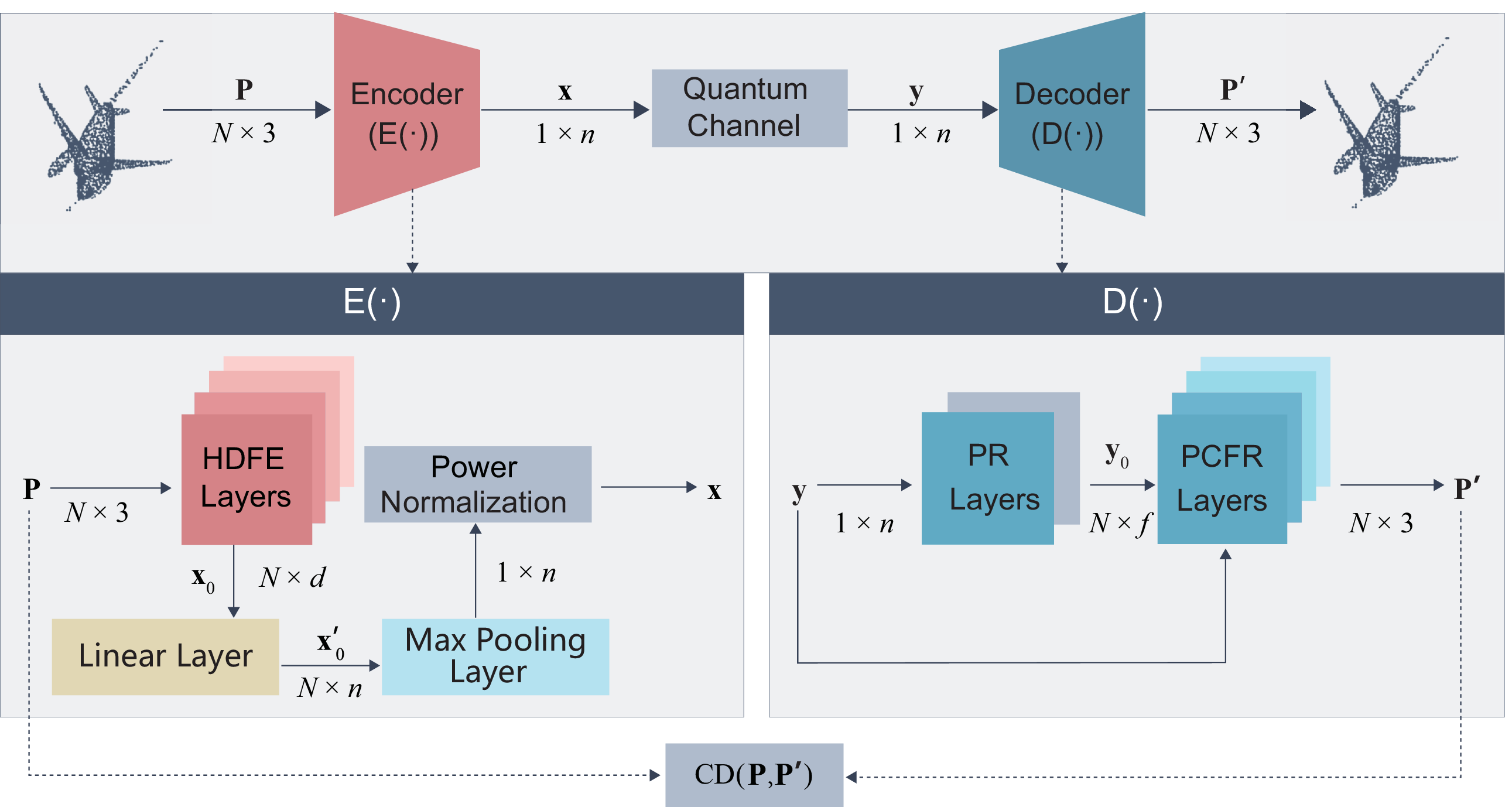}
	\caption{ Framework of the quantum semantic communication of point clouds.}
    \label{fig2}
\end{figure}

Our QSCPC system consists of two main components: an encoder network and a decoder network, as shown in the $\text E(\cdot )$ and $\text D(\cdot )$ parts of Fig.~\ref{fig2}. The point clouds $P$ contain three-dimensional coordinates, denoted as $\mathbf P\in \mathbb R^{N\times3}$ , where $N$ is the number of points.
 
 The encoder network takes a dense point cloud $\mathbf P$ as input and transforms it into point cloud semantic features $\mathbf x \in \mathbb{R}^{1 \times n}$, where $n$ denotes the length of transmitted data, a value that represents both the semantic code length and the raw data length in conventional transmission. To capture local geometric structures, a graph is first constructed from the input point cloud based on inter-point distances, using the $k$-Nearest Neighbors ($k$NN) strategy~\cite{Cover67}.
The point cloud $\mathbf P$ and the adjacency matrix of the graph are then input into five High-Dimensional Feature Extraction (HDFE) layers built upon graph convolutional networks~\cite{kipf2016semi}. Each HDFE layer gradually maps the point cloud features to higher-dimensional representations, thereby leveraging the local neighborhood correlations of the graph to deeply extract the semantic information of the point cloud. 
The high-dimensional features $\mathbf x_0 \in \mathbb R^{N\times d}$,  (where $d$  denotes the number of feature channels output by the last HDFE layer) are input into a linear layer.This layer maps $\mathbf{x}_0$ to a low-dimensional representation $\mathbf{x}_0^{'} \in \mathbb{R}^{N \times n}$. Next, a max-pooling layer selects the maximum value for each feature channel as the semantic information of the entire point cloud, thereby transforming the dimension of the point cloud features from $\mathbb R^{N\times n}$ to $\mathbb R^{1\times n}$. Finally, a power normalization operation is applied to obtain the point cloud semantic features $\mathbf x$, which are then transmitted through the quantum channel.

The semantic information of the point cloud $\mathbf x$ is directly encoded into the quantum states of the photons, and then transmitted with QSDC technology~\cite{Pan24,WangNSR25} over a noisy and high-loss channel to the receiver. We employ the STIKE one-way QSDC protocol~\cite{Pan25} to ensure secure and reliable transmission of semantic information. In the STIKE protocol, Alice encrypts information using preshared keys with the one-time-pad, and then transmitted using quantum states. Upon receiving, the receiver Bob decrypts the information and also distills a key from the transmitted quantum states simultaneously provided that the quantum bit error rate (QBER) is below the acceptable threshold. If there is no eavesdropping, the receiver recovers the point cloud semantic information $\mathbf x$.

The semantic features $\mathbf y$ received by the encoder network are obtained by transmitting $\mathbf x$ through a quantum channel, and $\mathbf y$ is used to obtain the reconstructed point cloud $\mathbf P^{'}\in\mathbb  R^{N\times 3}$. 
First, $\mathbf{y}$ is processed by the Point Restoration (PR) layers, each of which includes a 1D transposed convolution~\cite{Bian24}. Collectively, these two layers transform $\mathbf{y}\in \mathbb{R}^{1\times n}$ into the point cloud feature $\mathbf{y}_0 \in \mathbb{R}^{N\times f}$, where $f$ denotes the number of output feature channels.
Subsequently, $\mathbf y$ and $\mathbf y_0$  are input into the Point Cloud Feature Restoration (PCFR) layers. The first PCFR layer takes $\mathbf{y}_0$ as input, and each subsequent layer uses the output of the previous one as its input. In this way, the PCFR layers map features to different dimensions and guide their semantic fusion through $\mathbf{y}$, with each layer following the Point Condition Network (PCNet) architecture~\cite{Zheng24}. The final output feature dimension of the PCFR layer is 3, resulting in the reconstructed point cloud $\mathbf P^{'}\in \mathbb R^{N\times 3}$.

The system is trained using an end-to-end approach, with the objective of optimization of reducing the difference between the original point cloud distribution $\mathbf P$ and the reconstructed point cloud distribution $\mathbf P^{'}$. Chamfer Distance (CD)~\cite{Fan17} is a commonly used metric for measuring the difference between two point clouds by computing the bidirectional distance between them. It is computationally efficient and effectively captures the geometric discrepancy between the original and reconstructed point clouds, which makes it well-suited for use as both a loss function and an evaluation metric. The CD is defined as follows:
\begin{align}
\text{CD}(\mathbf P, \mathbf P^{'}) 
&= \frac{1}{|\mathbf P|} \sum_{a_i \in \mathbf P} \min_{b_j \in \mathbf P^{'}} \|a_i - b_j\|_2^2 \notag \\
&\quad + \frac{1}{|\mathbf P^{'}|} \sum_{b_j \in \mathbf P^{'}} \min_{a_i \in \mathbf P} \|b_j - a_i\|_2^2 .
\label{eq1}
\end{align}
where $a_i$ and $b_j$ denote the $i$-th and $j$-th points in  $\mathbf P$ and $\mathbf P'$, respectively, with $i,j = 1,2,\dots,N$.

\section*{Results}\label{sec: Results}
We used the ShapeNet\cite{Chang15} dataset in our experiments. The dataset contains 51,127 point cloud samples across 55 common object categories, such as airplanes, chairs, and beds. It was divided into 35,708 training samples, 5,158 validation samples, and 10,261 test samples. The compression rate is determined by the length of the semantic code $n$. For example, when $n=50$, the compression rate is $50/(2048\times 3) \times 100 \% = 0.81\%$. In this study, $n$ was set to $[10,20,50,100,200,300]$, and a separate model was trained for each case. Note that $n=6144$ represents the length of the raw point cloud when no semantic encoding is applied. Each model was trained for 200 epochs with a batch size of 32 using the Adam optimizer, with an initial learning rate of 0.001 that was halved every 20 epochs. After training, we used the models to encode and transmit test set point clouds. For each $n$, the corresponding model was evaluated, and the batch size was adjusted during transmission to match the bandwidth of the quantum channel.

\begin{figure*}[htbp]
	\centering
	\includegraphics[width=\textwidth]{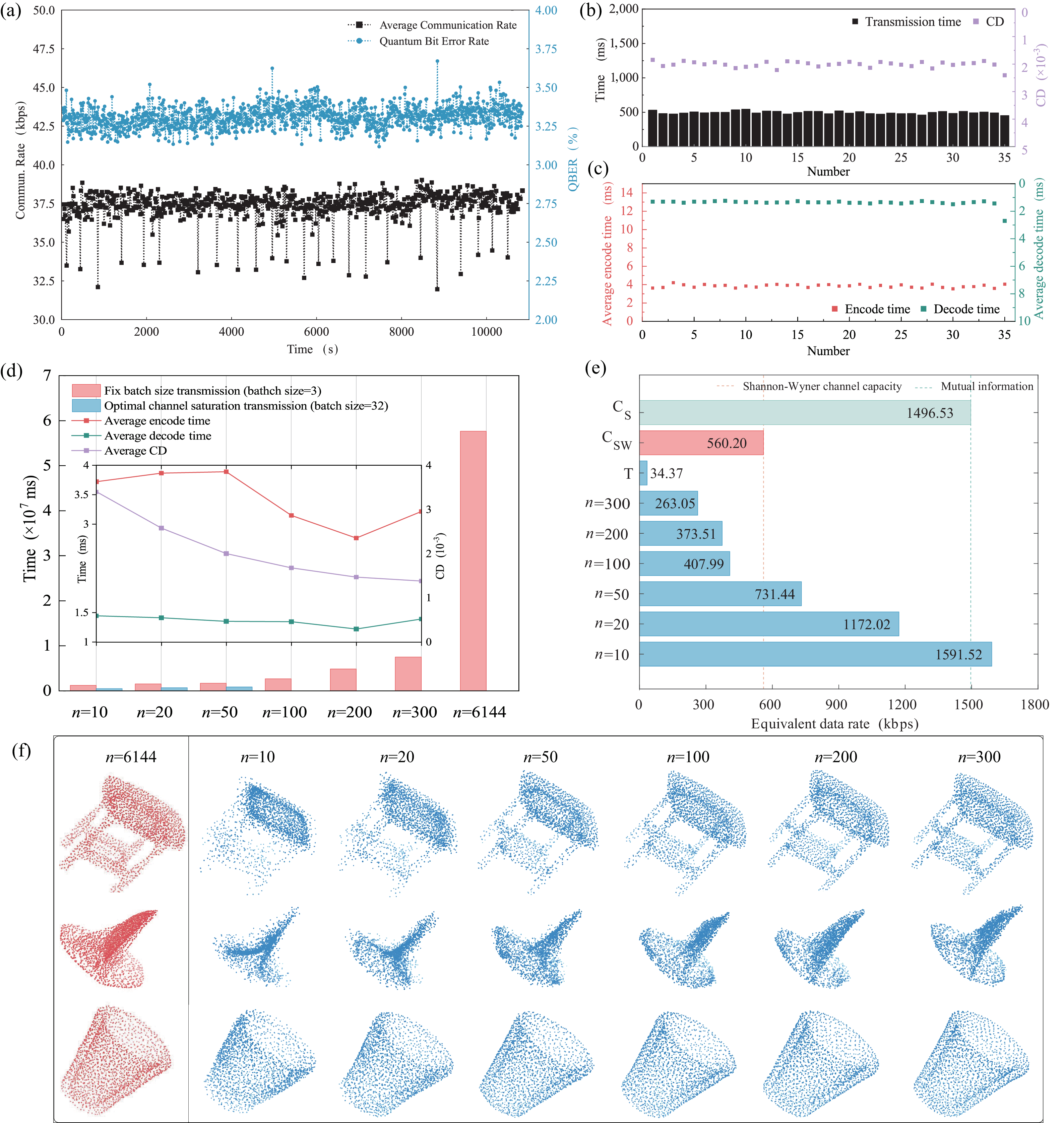}
	\caption{(a) Communication rate and QBER versus time of the QSDC system transmitting raw bits over a 50 km fiber link during 3 hours. 
    (b) Transmission time and CD under different Numbers when $n=50$. Each Number represents 100 transmission rounds. In each round, three point clouds are selected from the dataset, encoded into a $3\times n$ semantic feature, and transmitted once. The transmission time, measured from the start of encoding to the completion of decoding. 
    (c) Average encode time and average decode time under different Numbers when $n=50$. 
    (d) Comparison of transmission time, encode and decode, and CD under different $n$ and batch sizes. Transmission time denotes the total time to complete all transmission tasks corresponding to each $n$. For batch size 3, all $n$ are tested; for batch size 32, only $n=10,20,50$ are tested. In each transmission, a batch of point clouds is transmitted, and the process is repeated $10261/\text{batch size}$ times. Encode and decode times and CD values are computed for $n=10,20,50,100,200,300$. Note that $n=6144$ corresponds to transmitting the original point clouds without encoding.
    (e) Transmission efficiency for different $n$ with the Shannon–Wyner channel capacity as reference. At a fiber distance of 50 km, the first dashed line at 560.20 kbps corresponds to the secrecy capacity of the QSDC system derived from the Shannon-Wyner entropy, while the second dashed line at 1496.53 kbps marks the mutual information between the quantum transmitter and quantum receiver, namely the Shannon capacity. 
    (f) Visualization of reconstructed point clouds for different $n$.}
	\label{transmission}
\end{figure*}

For the transport layer, the transmitter and the receiver established a quantum channel via a 50 km standard single-mode fiber with an average link loss of 0.2 dB/km. The transmitter used weak coherent light as the light source, with an operating wavelength of 1550 nm, a repetition rate of 1.25 GHz, and a pulse width of 50 ps. The system employed one signal state and two decoy states, among which the average photon number of the signal state was 0.6, and those of the decoy states were 0.2 and 0 (vacuum state). To combat loss and noise in the quantum channel, coding techniques such as spread spectrum and low-density parity check (LDPC) have been utilized, resulting in a spreading ratio of 1:192 and forward error correction ratio of 1:12 in the experiment. The receiver adopted an InGaAs/InP single-photon detector, featuring a detection efficiency of 20\% and a dark count rate of $1.2 \times 10^{-6}$ per gate. As shown in Fig.~\ref{transmission}(a), the average quantum bit error rate (QBER) of the quasi-QSDC system was 3.31\%, far below the threshold of the one-way quasi-QSDC protocol~\cite{Pan25}, and the average communication rate during 3 hours of data collection was 37.36 kbps in a 50 km standard single-mode fiber link, thus guaranteeing safe and smooth transmission of semantic codes.

\textit{Transmission Data Analysis}: We first measured the encoding time, decoding time, CD, and the corresponding total transmission time at a fixed semantic code length of $n=50$. The batch size was set to 3 and the reported values are averages taken every 100 samples. As shown in Figs.~\ref{transmission}(b) and  (c), the encoding time is approximately 4 ms, the decoding time is approximately 1.5 ms, the channel transmission time is around 500 ms, and the CD fluctuates near $2\times10^{-3}$. These results indicate that the processing delays are negligible compared to channel latency, and that the system maintains a relatively low and stable distortion level. This demonstrates that semantic communication at $n=50$ can effectively balance transmission efficiency and reconstruction quality, which is crucial for real-time applications under bandwidth constraints.

We then evaluated the total transmission time of the test set, including both encode and decode and channel transmission. First, we compared the total transmission time for different code lengths under a fixed batch size of 3 and a saturated channel with a batch size of 32. As shown in Fig.~\ref{transmission}(d), using the saturated channel reduced the transmission time by 59.93\%, 54.48\%, and 47.51\% for $n=10$, $n=20$, and $n=50$, respectively, compared to the fixed batch size transmission. Furthermore, relative to the direct transmission of raw point clouds, the saturated channel reduced the transmission time by 91.34\%, 87.88\%, and 84.61\% for $n=10$, $n=20$, and $n=50$, respectively. These results clearly demonstrate the significant improvement of semantic communication in transmission efficiency under limited bandwidth.

We also investigated the impact of semantic code length on both processing delays and reconstruction accuracy. With a fixed batch size of 3, the average encoding and decoding times were evaluated at different values of $n$. The results are reported in The results are reported in Fig.~\ref{transmission}(d), where the maximum encoding time was 3.89 ms at $n=50$, while the maximum decoding time was 1.44 ms at $n=10$. Both delays remained at the millisecond level, satisfying the requirements for real-time communication. In addition, we analyzed the relationship between semantic code length and CD, where a CD value of 0 indicates that the reconstructed point cloud is identical to the original. CD decreased as $n$ increased, suggesting that larger code lengths enable the model to retain more geometric details and achieve higher reconstruction accuracy.

We finally summarize the transmission results in Table~\ref{tab1} to provide a more intuitive comparison. Specifically, we report the average encode and decode latency, the average CD values, and the total transmission time under different semantic code lengths $n$. To further evaluate the transmission performance, we introduce two metrics: Equivalent Data Rate (EDR) and Relative Transmission Efficiency (RTE). EDR is defined as the ratio between the total bits of all reconstructed point clouds and the total transmission time under a given code length $n$, while RTE is defined as the ratio between the transmission time of sending raw point clouds directly and that of tasks with code length $n$. Overall, we achieve an EDR of $34.37$ kbps when transmitting raw point clouds, while reaching $1591.52$ kbps at $n=10$, $731.44$ kbps at $n=50$, and $263.05$  kbps at $n=300$. As shown in Fig.~\ref{transmission}(e), the EDR at $n=50$ already exceeds the Shannon-Wyner channel capacity with a CD value of $2.00 \times 10^{-3}$, and the EDR at $n=10$ even exceeds the mutual information between the quantum transmitter and quantum receiver at the expense of data quality. In terms of RTE, the system achieves a $46.30$-fold efficiency improvement at $n=10$ compared to direct point cloud transmission. Considering the trade-off between quality and efficiency, a balanced performance can be obtained at $n=50$, which still provides a $21.28$-fold gain. Even when pursuing higher visual quality at $n=300$, the system maintains a $7.65$-fold improvement. 

\begin{table}[htbp]
    \caption{Performance of the Quantum Semantic Communication of Point Clouds}
    \centering
    \begin{tabular*}{\columnwidth}{@{\extracolsep\fill}ccccccc@{\extracolsep\fill}}
        \hline
        \multicolumn{1}{c}{$n$} & \multicolumn{3}{c}{Time (ms)} & \multicolumn{1}{c}{CD}& \multicolumn{1}{c}{EDR} & \multicolumn{1}{c}{RTE} \\
        \cline{2-4}
        & Encode & Decode & Total & ($\times 10^{-3}$) &(kbps)& \\
        \hline
        6144  & 0  & 0  & 57636000  & 0 &  34.37 & 1 \\
        10  & 3.72 & 1.44  & 1244715  & 3.40 & 1591.52 & 46.30 \\
        20  & 3.86 & 1.41  & 1690240  & 2.58 & 1172.02 &34.10 \\
        50  & 3.89 & 1.35  & 2708329  & 2.00 & 731.44 &21.28  \\
        100 & 3.15 & 1.34  & 4855498  & 1.68 & 407.99 &11.87 \\
        200 & 2.76 & 1.22  & 5303667  & 1.47 & 373.51 &10.87  \\
        300 & 3.21 & 1.39  & 7530870  & 1.38 & 263.05 &7.65  \\
        \hline
    \end{tabular*}
    \label{tab1}
\end{table}

\textit{Visualization of Transmission Results}:
To further illustrate the effect of semantic code length on reconstruction quality, we visualized the results for the Chair, Cap, and Mug categories at different values of $n$. As shown in Fig.~\ref{transmission}(f), larger code lengths produced smoother surfaces, more uniform point distributions, and more accurate geometry across the three categories. This confirms that increasing the semantic code length enhances reconstruction quality while maintaining transmission efficiency.

\section*{Discussion and Conclusion}

In this work, we propose a quantum semantic communication scheme, in which semantic communication improves efficiency by extracting essential information and eliminating redundancy, while quantum communication ensures that transmitted semantic features are safeguarded against potential adversaries. In particular, we experimentally demonstrate the quantum semantic communication of point clouds, and a 46.30-fold  efficiency improvement has been achieved at $n=10$ compared to direct point cloud transmission. Considering the trade-off between transmission quality and efficiency, a balanced performance can be obtained at $n=50$, where the equivalent data rate exceeds the Shannon-Wyner channel capacity with a CD value of $2.00 \times 10^{-3}$, and a 21.28-fold efficiency gain has been achieved. Even when pursuing higher visual quality at $n=300$, the system still provides a 7.65-fold  efficiency improvement.

Despite its potential, quantum semantic communication faces several challenges. Semantic communication models require further refinement to improve the generalization capability of models and to minimize semantic noise during the construction of semantic knowledge bases. Quantum channels are currently susceptible to environmental interference, and limited transmission distances necessitate the development of efficient quantum repeaters and error-correction schemes. As quantum memory, quantum error correction, and semantic communication protocols continue to mature, quantum semantic communication is expected to become a mainstream solution for smart cities, intelligent transportation, secure communications, and other sectors.

\section*{Acknowledgments}
This research was funded by the National Natural Science Foundation of China (62501057, 62471046 and 62131002), Young Elite Scientists Sponsorship Program by the China Association for Science and Technology (2022QNRC001), and Guangxi Science and Technology Program (2025GXNSFBA069227, GuikeAD23026225). 

\section*{Author contributions}
G.-L. L. proposed the protocol; L.-P. N. and G.-L. L. supervised the project; J. G. created the software; M. W., G.-F. Z. and G.-F. L. performed the experiments; M. W., G.-F. Z., G.-F. L. and G.-L. L. analyzed the data and wrote the manuscript; All authors revised the manuscript.

\section*{Competing interests}
The authors declare no conflict of interest.

\section*{Data availability}
Data are available from the corresponding author upon reasonable request.


\end{document}